\documentclass[aps,prl,reprint,amsmath,amssymb]{revtex4-2}
\usepackage{blindtext}
\usepackage{graphicx}
\usepackage{dcolumn}
\usepackage{bm}
\usepackage{amsmath}
\usepackage{braket}

\begin{document}

\title{Untangling Strongly and Weakly Interacting Configurations \\ in Many-electron Wave Functions}

\author{J. C. Greer}
\email{Jim.Greer@nottingham.edu.cn}
\affiliation{University of Nottingham Ningbo China\\
199 Taikang East Road, Ningbo, China 315100}
\altaffiliation{Nottingham Ningbo China Beacons of Excellence Research and Innovation Institute and the Department of Electrical and Electronic Engineering}

\date{\today}
 
\begin{abstract} 
Accurate solution of the many electron problem including correlations remains intractable except for few electron systems. Describing interacting electrons as a superposition of independent electron configurations results in an apparent combinatorial scaling to achieve an accurate solution. Many approximate approaches for large systems have been introduced, yet controlling and quantifying the error remains a significant hurdle for comparison to experiment. Perturbation calculations are a popular means to estimate neglected energy contributions but give rise to two unanswered questions. How to select a reference state ensuring perturbation corrections will be well-behaved? What is the residual error relative to the exact solution? Conditions to specify these two elusive requisites are given by introducing a partitioning of the electronic configurations into reference and interaction spaces with sampling of neglected configurations using renormalization group ideas. This approach is applied to the molecular (non-relativistic Coulomb) Hamiltonian to identify continuous phase transitions in the energy as a function of the number of configurations. Identification of the minimum number of reference configurations such that the neglected space consists primarily of independent energy contributions is achieved. Balancing the magnitude of the neglected contributions between two calculation enables prediction of accurate physical properties for which calculations would otherwise not be achievable using an expansion consisting of a full set of configurations. Consistent with the use of the molecular Hamiltonian, the analysis is valid for any finite fermion system described by a Hamiltonian consisting of one- and two-body operators.    
\end{abstract}

\maketitle

Solution of quantum many-electron problems underpins atomic, molecular and condensed matter physics, with similar methods applied in nuclear physics. The method of superposition of configurations, commonly known as configuration interaction (CI), remains a powerful method to accurately treat electron correlations but is not without limitations. The methods relies on a linear expansion of the many-electron wave function in determinants (or spin-projected determinants known as configuration state functions (CSFs)) constructed from spin orbitals as
\begin{equation}
\label{CI}
\ket{\Psi} = \sum_{i=1}^\infty c_i \ket{\psi_i},
\end{equation}
with $\ket{\Psi}$ the many-electron wave function, $\ket{\psi_i}$ a single determinant or CSF, with expansion coefficients $c_i$ determined from the variation principle. In the determinant basis, the Schr{\"o}dinger equation becomes a matrix eigenvalue problem
\begin{equation}
{\rm H} \, \vec{c} = E \, \vec{c},
\end{equation}
with energy matrix elements $H_{ij}=\bra{\psi_i}H\ket{\psi_j}$ and CI coefficient vector $\vec{c}$.
If $2m$ orthogonal spin orbitals are used to describe electrons with $n$ up and $n$ down spins, then  
\begin{equation}
\label{Mconfigs}
M = {m \choose n}^2
\end{equation}
determinants can be constructed. The total number of possible configurations $M$ can be reduced by introducing space and spin symmetries, nonetheless the underlying combinatorial nature of the expansion remains. Eq.~\ref{CI} becomes exact in the limit $M\rightarrow\infty$ whereas for computations $M$ is truncated. The solution for $M$ configurations built from a set of spin orbitals is referred to as the full CI (FCI) limit. Any practical attempt to work with the FCI space is limited to few electron problems. FCI calculations are nondeterministic polynomial-time complete (NP-complete) due to the scaling in the number of configurations thus remaining intractable for most systems. It is noteworthy that the electronic structure problem is believed to be QMA-complete, the quantum analog to NP-complete, suggesting obstacles for efficient quantum simulation algorithms of many-electron problems~\cite{OIW22}. 

Well-known since the early days of quantum theory is that for a FCI expansion, only a small fraction of the configurations contribute significantly to the wave function and energy. Means for truncating the CI expansion have been sought and in this context, selection methods~\cite{HMR73} were developed to exploit the structure of the CI matrix. The essential idea of selection methods is to start with a small number of configurations, generate interacting configurations relative to this set, retain important configurations based on selection criteria such as the contribution to the energy and wave function, and iterate to convergence~\cite{Har91,*Gre95,*TrF08,*BTA09,*LiH14,*Eva14,*HTU16,*OhH17,*GSG18,*PCP19,*TFL20,*ZLH21}. Selection methods remain among the most powerful techniques for solving many-electron problems~\cite{PCP23}. 

For a Hamiltonian with one- and- two-body potentials, energy matrix elements between determinants built from orthogonal electron states are zero unless the configurations differ in at most one or two orbitals. Assume that eq.~\ref{CI} is truncated and partitioned as
\begin{equation}
\label{partition1}
\ket{\Psi} = \sum_{v=1}^{N_v} c_v \ket{\psi_v} + \sum_{p=N_v+1}^{N_p} c_p \ket{\psi_p} 
+\sum_{n=N_v+N_p+1}^{N_n} c_n \ket{\psi_n},
\end{equation}
with the first sum $\ket{\Psi_v}$ treated variationally and defines a set of reference configurations, the second sum $\ket{\Psi_p}$ is treated by second order many-body corrections using Epstein-Nesbet perturbation theory (PT2) about the variational calculation, with the third sum representing neglected configurations $\ket{\Psi_n}$. Configurations to be calculated using the perturbation corrections are obtained relative to the reference set from
\begin{eqnarray}
\ket{\Psi_p} = 
\bigg( \sum_{\alpha;\iota} {\rm a}^\dagger_\alpha {\rm a}_\iota^{ } 
+ \sum_{\alpha,\beta;\iota,\kappa} {\rm a}^\dagger_\alpha {\rm a}^\dagger_\beta {\rm a}_\iota^{} {\rm a}_\kappa^{} \bigg)\ket{\Psi_v}. 
\end{eqnarray}
The action of the creation/destruction operators ${\rm a}^\dagger \, / \, {\rm a}$ creates a set of configurations differing in either one or two occupied orbitals relative to at least one configuration in the reference set. Here indices $\iota,\kappa$ refer to spin orbitals occupied in the reference whereas $\alpha,\beta$ are substitutions which lead to the generation of the perturbation set. Selecting $N_v$ fixes the number of configurations to be treated as perturbations $N_p$. For typical values of $N_v$, the increase in $N_p$ with $N_v$ is approximately linear. 

Under these assumptions, the matrix eigenvalue problem can be partitioned between a reference and perturbation space and written as
\begin{equation}
[{\rm H}_v+{\rm h}^\dagger (E-{\rm H}_p)^{-1}{\rm h}]\, \vec{c}_v \approx E \, \vec{c}_v.
\end{equation}
An additional condition is imposed requiring that the energy matrix elements between configurations in the perturbation set are non-interacting. This yields a diagonal matrix ${\rm H}_p \rightarrow {\rm D}_p$ with matrices ${\rm h}$ and ${\rm h}^\dagger$ composed solely of interactions between the reference and perturbation configurations. With this condition, the eigenvalue problem becomes  
\begin{equation}
\label{lowpart}
[{\rm H}_v+{\rm h}^\dagger(E^\prime-{\rm D}_{p})^{-1}{\rm h}]\, \vec{c}_v \approx E \, \vec{c}_v.   
\end{equation}
Both sides of the partitioned eigenvalue equation are energy dependent. If the equation is solved to self-consistency a Brillouin-Wigner perturbation expansion results, whereas the choices $E^\prime=E_v$ with 
\begin{equation}
E_v = \vec{c_v}^{\,\dagger} \, {\rm H}_v \, \vec{c}_v  
\end{equation}
together with intermediate normalization $\vec{c}_v^{\,\dagger}\, \vec{c}_v^{}=1$ leads to a Rayleigh-Schr{\"o}dinger perturbation expansion. For the latter case, the total energy is written as a sum of a variational reference energy and a sum of non-interacting perturbations as
\begin{equation}
\label{varpert}
E \approx \vec{c}_v^{\,\dagger} \,{\rm H}_v\, \vec{c}_v + \, 
          \sum_{p=1}^{N_p} \frac{ | \bra{\Psi_v}  H \ket{\psi_p} |^2 } {E - H_{pp} }.
\end{equation}
The partitioning of a CI matrix into a strongly interacting reference with weak interactions to all other configurations is at the heart of the distinction between static and dynamic correlations, respectively.

In contrast to selection methods which start with a small set of configurations and increase the expansion, eq.~\ref{lowpart} can be also considered as an approximation to the FCI wave function as the number of configurations is reduced. Let the reference set be chosen as $N_v=M-1$ with the ordering of the configurations corresponding to their contribution to the energy leading to 
\begin{equation}
\label{iter1}
E  \approx \vec{c}_v^{\,\dagger} \,{\rm H}_v\, \vec{c}_v |_{M-1} 
+\frac{ | \bra{\psi_M}  H \ket{\Psi_{v;M-1}} |^2 } { E - H_{M\,M}}.    
\end{equation}
Next the reference set is chosen such that the $M^{th}$ and $(M-1)^{th}$ configurations are treated as perturbations with the assumption that $\bra{\psi_{M-1}} H \ket{\psi_M}=0$, leading to
\begin{equation}
\label{iter2}
E  \approx \vec{c}_v^{\,\dagger} \,{\rm H}_v\, \vec{c}_v |_{M-2}
+\sum_{p=M-1}^M\frac{  | \bra{\psi_p}  H \ket{\Psi_{v;M-2}} |^2 } { E - H_{p\,p}}.
\end{equation}
Iterating in this way leads to eq.~\ref{varpert}. As configurations are moved from the variational set, the change in the variational energy ideally should change by the same magnitude but with opposing sign as the PT2 correction. However, the PT2 correction does not allow for changes to the reference wave function so the lack of relaxation to the reference set introduces a small error even at larger values of $N_v$.  As $N_p$ increases, $N_v$ decreases and the reference set will increasingly become dominated by lower order excitations. The reference and higher order excitations decouple and do not contribute to eq.~\ref{varpert}. As configurations become neglected, the energy error should be maintained to within an acceptable level. The two key conditions for maintaining an acceptable error are that the perturbations are independent of one another and  that $N_v$ is large enough such that there are enough $N_p$ corrections to approximate the FCI energy. The latter condition is equivalent to the statement that the $N_v$ should not be reduced below a value for which the neglected configurations can be considered negligible. 

\begin{figure}
\begin{center}  
   \centering
    \includegraphics[width=1\columnwidth]{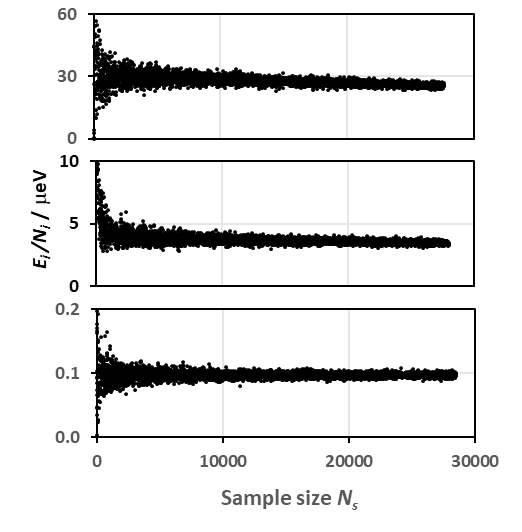}
    \caption{\label{fig1} Distribution of sampled average energies relative to a fixed vector length $N_v$. Upper panel: $N_v$ = 931; Middle panel: $N_v$ = 4,182; Lower panel: $N_v$ = 44,981}
\end{center}
\end{figure}

To explore when these conditions hold, the ground state energy of molecular nitrogen N$_2$ is investigated using the Monte Carlo configuration interaction method (MCCI)~\cite{Gre98,CoP12}. Hartree-Fock orbitals are calculated~\cite{GAMESS} using an augmented triple $\zeta$ (aug-cc-pVTZ) Gaussian basis~\cite{KDH92}. The orbitals are used to generate the one- and two-electron integrals for calculation of the CI matrix elements. In the CI calculations, the core molecular orbitals dominated by $1s$ core states are frozen. The initial MCCI trial vector is often chosen as the lowest energy configuration. Due the stochastic selection scheme, the choice of initial configuration is not critical but ideally belongs to the reference set. To expand the CI vector, a branching step consisting of randomly selecting single and double excitations is performed. With the new vector, the CI matrix is diagonalized and a pruning step is next performed consisting of discarding configurations with coefficient magnitudes less than a threshold tolerance. After the branching and pruning steps, the number of new configurations and the change in energy are checked. If the calculation has not converged, a new random generation of single and double excitations relative to the previous iteration's CI vector is performed, with the steps continued to convergence~\cite{Gre98}. Typically for selection methods, a rapid decrease followed by a slow decrease in energy is seen as the number of configurations is increased indicating strongly interacting configurations are readily found whereas there are many more configurations with small contributions. Weakly interacting configurations can be treated by PT2 corrections to the variational calculation and likewise the reference set can be expanded if important configurations are identified from the PT2 corrections~\cite{CoP12}. For this study, standard MCCI calculations are performed for different convergence criteria leading to a set of increasingly accurate CI vectors with differing lengths. Relative to each of the vectors, 5000 random samples are generated with between 1 $\leq N_i \leq$ 30,000 additional interacting configurations. Histograms for the change in the CI energy per configuration 
\begin{equation}
E_i/N_i=[E(N_v)-E(N_i)]/N_i
\end{equation}
due to the addition of sampled configurations are shown within fig.~\ref{fig1} and note, for this definition, the change in energy is positive with increasing correlations. $E(N_v)$ is the variational energy without sampled configurations and $E(N_i)$ is the variational energy with $N_i$ added configurations. An average energy density for the samples is defined as
\begin{equation}
\label{expect}
\langle E/N \rangle = \frac{1}{N_s} \sum_{i=1}^{N_s} E_i/N_i
\end{equation}
with $N_s=5000$.  

\begin{figure}
\begin{center}  
    \centering\includegraphics[width=1\columnwidth]{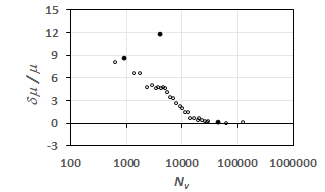}
    \caption{\label{fig2} The scaled energy difference $\delta \mu$ give by eq.~\ref{dmu}. The larger, filled circles correspond to the values at $N_v$=931 ; 4,182 ; 44,981 for comparison to fig.~\ref{fig1}.}
\end{center}
\end{figure}

In fig.~\ref{fig1}, a large skew in the histograms can be observed for the smaller reference vectors in contradiction to the requirement that the perturbation corrections be independent which would lead to constant average energy density. As $N_v$ is increased, the skew reduces significantly in the lower panel of fig.~\ref{fig1}. It is noted that both the energy and the number of configurations fluctuate during sampling. Maximizing the probability entropy to determine the most likely distribution for the sampled set yields a grand canonical distribution, from which the derivative of the average energy with respect to $N_i$ may be written as
\begin{equation}
d\langle E \rangle /dN 
= \frac{\langle EN\rangle -\langle E\rangle \langle N\rangle}
{ \langle N^2\rangle - \langle N \rangle^2 } 
= \frac{{\rm covar}(EN)}{\rm{var}(N)} .
\end{equation}
The probability for each sample is evaluated for $p_i=1/N_s$ as in eq.~\ref{expect} corresponding to the 'high temperature' limit for the statistical averages. If the energy density for each sample becomes constant for sampling relative to a fixed variational vector or  
\begin{equation}
\label{chempot}
E_i =\mu(N_v) \, N_i,
\end{equation}
then the condition
\begin{equation}
\label{noskew}
\langle E/N \rangle = d\langle E \rangle /dN = \mu(N_v)    
\end{equation}
holds and $\mu(N_v)$ can be viewed as the chemical potential for the interacting configurations. In this case, the average energy is independent of the configurations that have been randomly selected such that eq.~\ref{lowpart} becomes a valid approximation. In fig.~\ref{fig2}, the scaled energy difference 
\begin{equation}
\label{dmu}
\delta \mu/\mu =\big[ \langle E/N \rangle - d\langle E \rangle/dN \big] / \langle E/N \rangle
\end{equation}
is shown. For large values of $N_v$, there is very little skew in the energy with sample size corresponding to $\delta\mu / \mu\approx 0$. As $N_v$ is reduced, the skew in $E(N_i)$ increases until changing abruptly in the vicinity of $N_v=4182$.

\begin{figure}
\includegraphics[width=1\columnwidth]{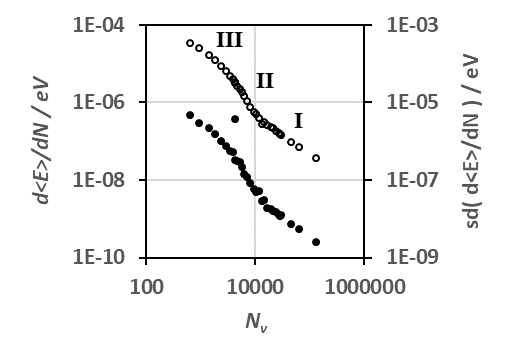}
\caption{\label{fig3} The upper curve shows the change in sampled energy with sample size $\langle E\rangle /dN$ and the lower curve is the standard deviation $sd(\langle E\rangle/dN$).}
\end{figure}

To explore the interacting space, $d\langle E\rangle/dN$ and its standard deviation are plotted in fig.~\ref{fig3} as a function of $N_v$. Three regions labeled by I, II, and III can be seen. As $N_v$ decreases and $d\langle E\rangle/dN$ begins to increase, the standard deviation begins changing in a disordered way. The small difference in the change in energy per configuration between region I to II results in the disorder due to mixing from configurations in the two regions near the transition. As $N_v$ further decreases, a sharper transition between regions II and III occurs indicating a much larger change in energy per configuration between these two regions. The values of $N_v$ in region II are too small to include all strongly interacting configurations needed in the reference. This leads to samples with varying numbers of strongly and weakly interacting configurations, whereas in region I the samples are dominated by weakly interacting configurations. Region III is defined by values of $N_v$ that are so small that the interacting space is dominated by configurations belonging to the reference set. Iterating the partitioned Hamiltonian as in eqn.~\ref{iter1} and \ref{iter2} in a renormalization group fashion with a stochastic sampling of the interacting configurations leads to a mapping of the strength of electron correlations as a function of $N_v$. This sets a lower limit for $N_v$ which for which PT2 corrections are valid. Below this limit, the increasing skew in the sampled energies signals a larger reference set is needed. 

\begin{figure}
\centering
    \centering
    \includegraphics[width=\columnwidth]{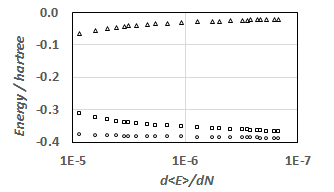}
    \caption{\label{fig4} Energy convergence as a function of $d\langle E\rangle /dN$. Squares - variational energy. Triangles - perturbation correction. Circles - variational energy plus perturbation correction.}
\end{figure}

In fig.~\ref{fig4} the variational, PT2, and total energy estimate energies as functions of $d\langle E \rangle /dN$ are shown. To estimate the error with respect to the FCI limit, the sum $E_v+E_p$ for values satisfying eq.~\ref{chempot} can be extrapolated against $d\langle E \rangle /dN$ or $N_p$. This is one approach to the error estimate, machine learning~\cite{CaT17,Coe18} and eigenvector continuation~\cite{FHI18,SaL21} methods that extrapolate correlation energies at low computational cost are being developed. However, simple extrapolation leads to an estimate of the FCI energy of -109.3748$\pm$0.00033 hartree and -109.3745 hartree using $d\langle E \rangle /dN$ and $N_p$, respectively, as independent variables. The linear regression as a function of $d\langle E \rangle/dN$ takes into account the sampling and convergence errors providing a standard error~\cite{YEM04}. The FCI estimate is consistent with the value of -109.3747$\pm$0.0003 hartree obtained in ref.~\cite{Fel99}, although a different bond length of 110.040 pm and the cc-pVTZ orbital basis was used compared to 109.434 pm and the aug-cc-pVTZ basis, respectively, used in this work. This can be compared to a recent blind benchmark test for several methods on the benzene molecule using a correlation consistent double $\zeta$ (cc-pVDZ) Gaussian basis~\cite{Dun89}. A key finding in the study is a root-mean-square deviation in the predicted energies across a variety of configuration based methods is 0.0013 (0.035) hartree (eV)~\cite{BzGS}. In the present work, the largest explicit calculation is with $N_v=28,963$ and $N_p \approx 2.305\times 10^8$ with the interacting set increasing at most linearly with $N_v$ (for comparison, from eq.~\ref{Mconfigs} the naive estimate of the total number of configurations is approximately $1.9\times 10^{15}$). The resulting FCI energy estimate is $E_v+E_p$ = -109.3735 hartree. This leads to an estimated error relative to the FCI estimate in the range of approximately 0.00097 (0.026) and 0.00016 (0.044) hartree (eV). These errors can be significant for quantities such as transition barriers and dissociation energies and both quantities rely on energy differences. Hence determining and balancing the error between two calculations due to neglected configurations remains key to determining physical quantities. 

A procedure for partitioning the many-electron problem emerges by characterizing interacting configurations relative to a reference state obtained from a selection method. In the context of historical discussions on the division between static and dynamic correlations, the size of the reference set as defined is found to be much larger than typically assumed. The goal of the partitioning is to avoid the need for FCI calculations while avoiding combinatorial scaling. It also provides a framework for making energy estimates for determination of the relative error between calculations. The iterative variational (matrix diagonalization) calculations scale as order ${\cal O}(N_v^2)$ and the perturbation calculation can be reduced to ${\cal O} (N_p)$, or more commonly as ${\cal O}(N_p \ln N_p)$ including the removal of duplicate configurations during generation of interacting configurations. A sequence of reference sets of increasing size can be selected. Adding interacting configurations acting as independent contributions relative to the reference sequence, an estimated error relative to the FCI energy can be extracted. In practical calculations, sampling $d\langle E \rangle / dN$  can confirm if perturbation corrections will be accurate and measure the level of convergence of a selection method. The goal for partitioning the many electron problem is to overcomes the CI scaling problem. Mapping electron correlations through the selection of dominant reference configurations together with sampling of the interacting space provides a measure of the accuracy of a many-electron calculation for individual systems, and provides a means to balance errors between two systems.

\begin{acknowledgments}
This work has been supported by the Nottingham Ningbo China Beacons of Excellence Research and Innovation Institute. 
\end{acknowledgments}

\bibliographystyle{apsrev4-2} 
\bibliography{main} 

\end{document}